\documentclass[aps,prb,twocolumn,showpacs,groupedaddress]{revtex4}

\def\today{30 Sept. 2013}

\usepackage{amssymb}

\usepackage{graphicx}
\usepackage{bm}

\begin{document}

\title{

\begin{minipage}[t]{7.0in}
\scriptsize
\begin{quote}
\leftline{
{\it Phys. Rev. B}, in press.
}
\raggedleft {\rm arXiv:1307.6260}
\end{quote}
\end{minipage}

Resistance asymmetry of a two-dimensional electron gas caused by an 
effective spin injection
}
\author{D. I. Golosov}
\email{Denis.Golosov@biu.ac.il}
\affiliation{Jack and Pearl Resnick Institute of Advanced Technology, Department of Physics, Bar-Ilan University, Ramat-Gan 52900, Israel}
\author{I. Shlimak}
\affiliation{Jack and Pearl Resnick Institute of Advanced Technology, 
	Department of Physics, Bar-Ilan University, Ramat-Gan 52900, Israel}
\author{A. Butenko}
\affiliation{Jack and Pearl Resnick Institute of Advanced Technology, 
	Department of Physics, Bar-Ilan University, Ramat-Gan 52900, Israel}
\author{K.-J. Friedland}
\affiliation{Paul-Drude Institut f\"{u}r Festk\"{o}rperelektronik, Hausvogteiplatz 5-7,
10117, Berlin, Germany}
\author{S.~V. Kravchenko}
\affiliation{Physics Department, Northeastern University, Boston, Massachusetts 02115,
U.S.A.}
\date{\today}

\begin{abstract}
We have performed conductivity measurements on a Si-MOSFET sample 
with a slot in the upper gate, allowing for different electron 
densities $n_{1}$ and $n_{2}$ across the slot. Dynamic longitudinal 
resistance was measured by a standard lock-in technique, while 
maintaining a large DC current through the source-drain channel.  
We find that in a parallel magnetic field, the resistance of the sample, 
$R(I_{\mathrm{DC}})$,  is asymmetric 
with respect to the direction of the DC current. The asymmetry becomes
stronger with an
increase of either the magnetic field or  the difference between $n_{1}$ and 
$n_{2}$. 
These observations are interpreted in terms of the effective spin injection:
the degree of spin polarisation is different in the two parts of the sample, 
implying different magnitudes of spin current away from the slot.  
The carriers thus leave the excess spin (of the
appropriate sign) in the region around the slot, leading to spin accumulation
(or depletion) and to the spin drift-diffusion phenomena. Due to the
positive magnetoresistance of the two-dimensional electron gas, this
change in a local magnetisation affects the resistivity near the
slot and the measured  net resistance, giving rise to an asymmetric 
contribution. 
We further observe that the value of $R(I_{\mathrm{DC}})$ saturates at large 
$I_{\mathrm{DC}}$; we suggest that this is due to electron tunnelling
from the two-dimensional n-type layer into the p-type silicon (or into
another ``spin reservoir'') at the
slot.
\end{abstract}

\maketitle

\section{Introduction}

The objective of this work was to probe the influence of strong 
parallel magnetic field on the electron transport across an interface 
between regions with different electron densities $n_{1}$ and $n_{2}$ 
in a single Si-MOSFET sample. The sample has a narrow slot of 90 nm 
in the upper gate, which allows to apply different voltages to separate 
gates. Previously, longitudinal conductivity of a slot-gate Si-MOSFET 
sample was measured in a perpendicular magnetic field, in the quantum 
Hall effect (QHE) regime~\cite{Shlim}. It was shown that for 
sufficiently large electron concentrations on the two sides of the narrow slot,
 the presence of the slot does not give rise to 
a measurable resistance increase. This 
implies that the slot does not act as a potential barrier for electrons.
  
The effect of a parallel magnetic field on the conductance of a 
two-dimensional electron gas (2DEG) in spatially uniform Si-MOSFET 
samples has been investigated earlier~\cite{Shashkin1, Simonian, Tsui} in the context of 
metal-insulator transition studies. The conductance asymmetry with 
respect to the direction of the electric current (parallel or antiparallel 
to the magnetic field), reported here, is a novel effect associated 
with the non-uniform properties of our slot-gate sample. 
Phenomenological interpretation of our results (involving 
current-induced spin accumulation or depletion near the slot) suggests 
that this asymmetry is directly related to the physical mechanism 
underlying the positive magnetoresistance of a Si-MOSFET in parallel 
magnetic fields\cite{Shashkin1}.

When a uniform 2DEG is placed in a parallel magnetic field, applying 
a source-drain voltage gives rise to both charge and spin currents, and the
ratio of the two depends on the carrier spin polarisation and therefore on 
the carrier density. In our case, two 2DEG
systems of different densities are connected in  series (by the region
underlying the slot in the gate). Then
the magnitudes of spin current far away from the slot 
(where the system can be viewed as uniform) are different on the two sides
of the slot.
Therefore, the carriers leave the excess spin (of the appropriate sign
depending on the direction of the electrical current) in the region of the 
slot,
giving rise to the {\it effective spin injection} (cf. Ref. 
\onlinecite{injection}) . This results in changing
the net carrier spin in the vicinity of the slot.
The latter in turn affects the resistivity of the 2DEG, and thus the 
conventional resistance
measurements contain information about the local carrier spin polarisation. 
The sign of the measured correction to the dynamic resistance
depends on whether the carrier spin is accumulated or depleted (i.e., on the
sign of the current), hence the observed resistance asymmetry. Thus, in
our experiment we  perform the effective 
spin injection while also measuring its rate.

With increasing electrical current, the asymmetric contribution to the 
resistance appears to saturate. We suggest that
this is a consequence of spin current ``leakage'' at the slot, due to the
tunnelling into, {\it e. g.,} the underlying p-type silicon. With increasing
DC current,
spin accumulation or depletion in the slot region become more pronounced.
This, in turn, leads to an increased rate of the ``leakage'', thereby 
restricting further increase of spin accumulation/depletion 
and that of the associated resistance asymmetry.

The paper is organised as follows: after describing the experimental
procedure in Sec. \ref{sec:exp}, we give an overview of the data 
and summarise the basic theoretical ideas in Sec. \ref{sec:overview}.
This is followed by a more detailed discussion of the theoretical model
(Sec. \ref{sec:theory}), and a comparison with experimental results 
is found in Sec. \ref{sec:compare}.

Preliminary results were published in Ref. \onlinecite{Shlimak09}.

\section{Experimental}
\label{sec:exp}

\begin{figure}
\includegraphics[width=8.5cm]{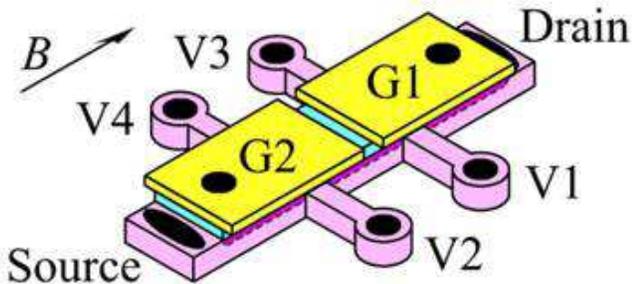}
\caption{\label{fig:scheme} (colour online). Schematic view of the sample.}
\end{figure}

The sample used in our experiments (see Fig. \ref{fig:scheme})  was studied earlier in Ref. \onlinecite{Shlim}.
The width of the 2DEG channel is 30 $\mu$m. 
Narrow slot ($\sim$~90~nm) was made in the upper metallic gate, 
allowing to apply different gate voltages to different parts of the 
gate and thereby to independently control the electron density in the 
two areas of the sample. The distance between the slot and the contact V1 (V2) is 30 $\mu$m (150 $\mu$m).
 By measuring the transverse Hall resistivity, 
$\rho_{xy}$, and longitudinal resistivity, $\rho_{xx}$, as functions 
of the gate voltage $U_\mathrm{G}$ in a perpendicular magnetic field 
we obtained the dependence of electron density $n$ on $ U_\mathrm{G}: 
n = 1.43\cdot 10^{15}\cdot(U_\mathrm{G}-0.64V) \mathrm{m}^{-2}$, with electron 
mobility 1.46~m$^2$/V$\cdot$s at $n = 1.62\cdot 10^{16} \mathrm{m}^{-2}$.

For the present experiment, the sample was mounted along the 
magnet axis, so that the current flow would be parallel to the magnetic 
field. The misalignment between the two was estimated with the help of 
Hall effect measurements. Whereas the Hall voltage must vanish for the 
ideal planar geometry, the small value registered corresponds to a 
minute out-of-plane misalignment of about  $\sim 0.1^\circ$.

Our experimental scheme enables one to pass a 
large DC current, 
$I_\mathrm{DC}$, of about 1~$\mu$A through the source-drain channel, while 
measuring the dynamic resistance at 12.7~Hz frequency by means of a 
standard lock-in technique with an AC current of 10--50~nA. Sample 
temperature was maintained at 0.3~K.
  
\begin{figure*}
\includegraphics[width=16.5cm]{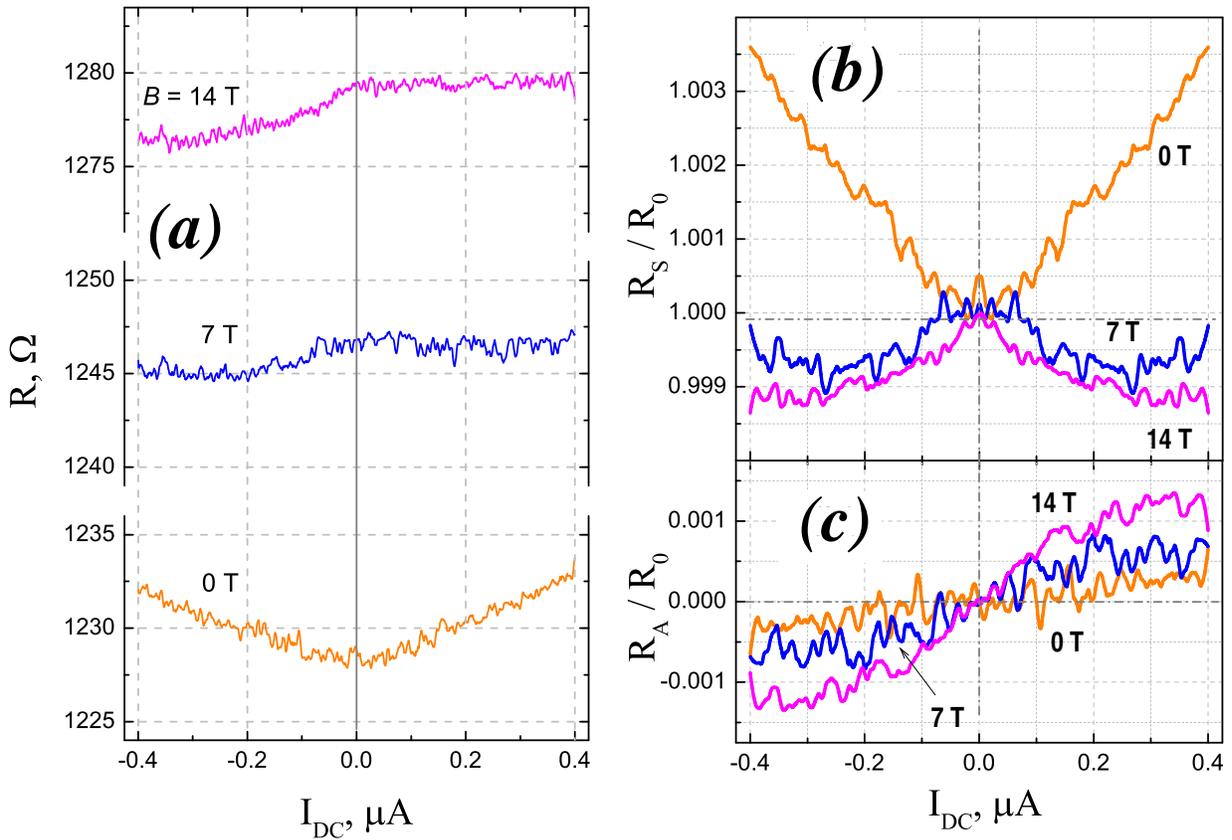}
\caption{\label{fig:datafield} (colour online).
(a) Dynamic resistance as a function of DC current at $B = 0$, $7$, 
and  
$\pm14$~T . $U_{\rm G}(1) = 7$~V, $U_{\rm G}(2) = 18$~V, corresponding
to carrier densities $n_1=0.9 \cdot 10^{16}$ m$^{-2}$ and 
$n_2=2.5 \cdot 10^{16}$   m$^{-2}$.
Panels (b) and (c)  show normalised symmetric  and
antisymmetric  parts of the data shown in (a).}
\end{figure*}

In the first series of measurements, we fix different gate voltages 
applied to the different areas of the sample across the slot: in area 
1, $U_\mathrm{G}(1) = 7 V$, which corresponds to 
$n_1 = 0.9\cdot 10^{16} {\rm m}^{-2}$, and in area 2, 
$U_\mathrm{G}(2) = 18 V$, which corresponds to 
$n_2 = 2.5\cdot 10^{16} {\rm m}^{-2}$. Then we measure the dynamic 
resistance of the sample as a function of DC current at zero magnetic 
field and in 
parallel fields $B = 7$ and 14 Tesla (Fig.~\ref{fig:datafield}). 
One can see the 
following features:

\noindent (i) At zero $I_\mathrm{DC}$, a positive 
magnetoresistance\cite{Shashkin1,Simonian,Tsui,Shashkin,Okamoto5,Vitkalov,Lai,Okamoto8,Shlimak12} (PMR) 
is observed: resistance increases with  magnetic field. 

\noindent (ii) 	At $B=0$, resistivity slightly increases with the DC current, 
and 
$R(I_\mathrm{DC})$ is almost symmetric with respect 
to the direction of $I_\mathrm{DC}$.

\noindent (iii)	At $B = 7$ and 14~T, the dependencies $R(I_\mathrm{DC})$ are 
clearly asymmetric. This asymmetry, which increases with $|B|$, does not 
depend on the direction of 
the magnetic field: the shape of the curves is identical for $B = 14$~T 
and $B = - 14$~T. This excludes Hall voltage (which may arise due to a 
slight misalignment of the sample) as a possible origin of the asymmetry.
  
\begin{figure}
\includegraphics[width=8.5cm]{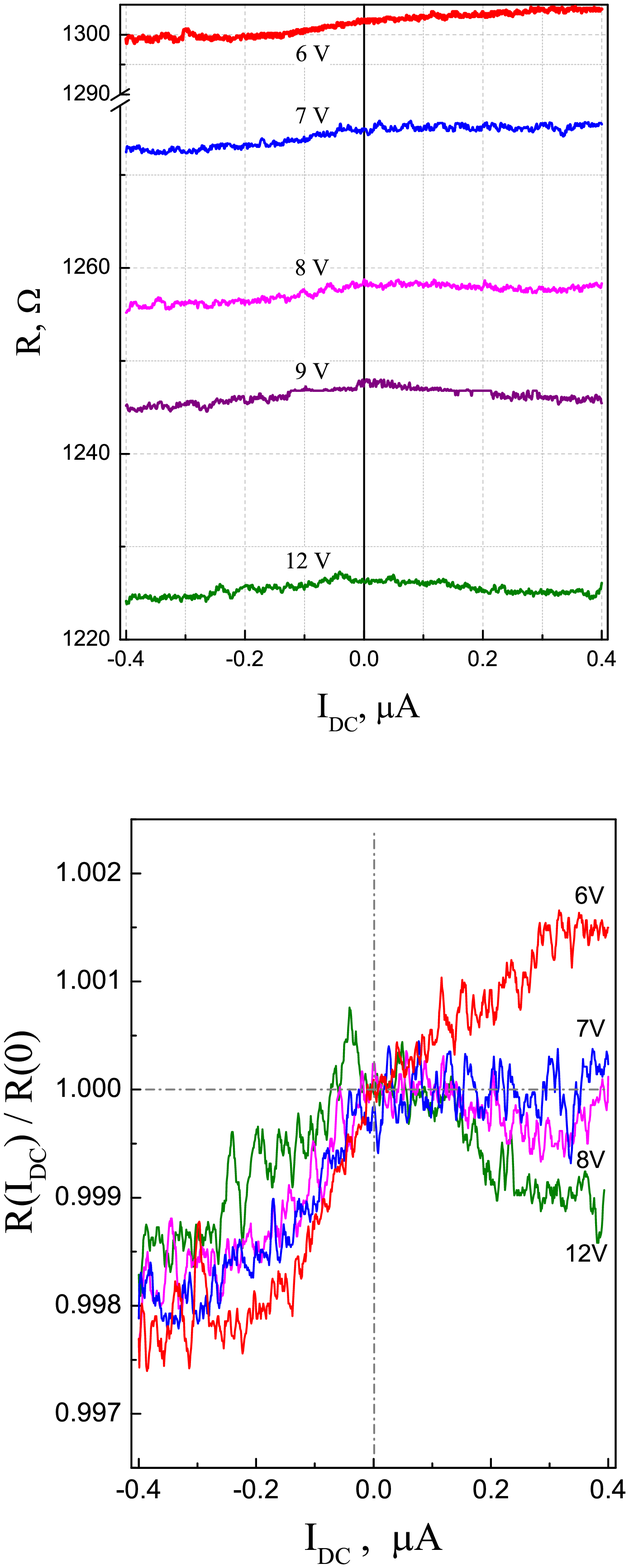}
\caption{\label{fig:datagate} (colour online).
Top panel: dynamic resistance as a function of DC current for varying values
of the gate voltage $U_G(1)=6,7,8,9$ and $12$V (corresponding 
to $n_1= 0.83\cdot 10^{16},  \,0.95\cdot 10^{16},  \,1.2\cdot 10^{16},
\,1.25\cdot 10^{16} $ and $1.7\cdot 10^{16}$ m$^{-2}$). $U_G(2)$ is fixed at 
$18$ V ($n_2=2.5 \cdot 10^{16}$ m$^{-2}$),
and $B=14$T. The systematic relative increase of
the resistance asymmetry with increasing $n_2-n_1$ is highlighted by the bottom panel.
}
\end{figure}

  In the second series of measurements, dependencies 
$\rho(I_\mathrm{DC})$ were obtained at $B = 14$~T for the case when 
one gate voltage was maintained at a constant value, 
$U_\mathrm{G}(2) = 18 {\rm V}$, while the other 
varied from $U_\mathrm{G}(1)=12 {\rm V}$ to  $U_\mathrm{G}(1)=6 {\rm V}$ 
( Fig.~\ref{fig:datagate}). 
One can see that asymmetry increases with the increase of the difference 
between $n_1$ and $n_2$. 

\section{Experimental Results and Basic Interpretation -- an Overview}
\label{sec:overview}

   Positive magnetoresistance (PMR)  effect in parallel magnetic fields in 
Si-based 
two-dimensional systems has been observed 
earlier\cite{Shashkin1,Simonian,Tsui,Shashkin,Okamoto5,Vitkalov,Lai,Okamoto8,Shlimak12}. 
It was shown in Refs. \onlinecite{Shashkin,Okamoto5} that the 
metallic-like conductivity of Si 
MOSFET first decreases with increase of in-plane magnetic field and 
then saturates  to a new constant value when electrons become fully 
polarised. This effect is variously attributed to the reduction of screening 
of 
charge impurities in a Fermi liquid caused by the loss of spin 
degeneracy\cite{Dolgopolov}, or to a combined effect of spin polarisation,
interaction, and multiple impurity-scattering\cite{Zala}. The reader
is referred to Ref. \onlinecite{Kuntsevich} for further discussion.
What is important for us presently is that the PMR effect is of spin 
origin, {\it i. e.,} the conductivity depends on spin polarisation (or
equivalently, on magnetisation), which
in turn is affected by the applied magnetic field.

  In the present paper, we are interested in the asymmetry of the
measured resistance $R(I_\mathrm{DC})$ with respect to the sign of 
$I_\mathrm{DC}$ . Figs.~\ref{fig:datafield}  (b) and (c) show the result of 
decomposition of 
$R(I_\mathrm{DC})$ into symmetric $R_\mathrm{S}$ and 
antisymmetric $R_\mathrm{A}$ parts: 
$R_\mathrm{S}(I_{\rm DC})=[R(I_{\rm DC})+R(-I_{\rm DC})]/2$, and 
 $R_\mathrm{A}(I_{\rm DC})=R(I_{\rm DC}) - R_\mathrm{S}(I_{\rm DC})$. For 
convenience, we show normalised values,
$R_\mathrm{S,A}/R(I_\mathrm{DC}=0)$.
The profile of $R(I_\mathrm{DC})$ 
at $B = 0$ is almost symmetric. This symmetric increase is presumably due
to the conductivity being strongly affected by the Joule heating
(of the electron system), 
proportional to $(I_\mathrm{DC})^2$. In our case, both electron 
concentrations $n_1$ and $n_2$ correspond to the metallic side of the 
metal-insulator transition in 2D electron systems, when 
$d R /dT > 0$, so increasing the temperature must lead to a 
resistance increase, explaining the experimental observation. 
The small 
asymmetry observed at $B = 0$ (about 2.5$\cdot 10^{-4}$ of the net 
resistance at maximal current) can be explained by an additional 
voltage bias $V_\mathrm{DC}$ induced by the DC current: 
$V_\mathrm{DC} = I_\mathrm{DC} R$. In MOSFETs, 
$V_\mathrm{DC}$ is added to the gate voltage $U_\mathrm{G}$ with an 
appropriate sign (cf. ``pinch-off'' effect \cite{Cobbold}). 
For our sample geometry, $V_\mathrm{DC}$ at 
$I_\mathrm{DC} = 0.4 \mu \mathrm{A}$ reaches 1~mV which is, indeed, 
about 10$^{-4}$ of the $U_\mathrm{G}$. This leads to a small 
increase or decrease (depending on the sign of $I_\mathrm{DC}$) of 
the electron density and corresponding asymmetric contribution to 
the sample resistance.

  It follows from Fig. \ref{fig:datafield}, that in strong parallel magnetic 
fields 
($B = 7$ and 14~T), the Joule heating due do the DC current $I_\mathrm{DC}$ 
does not influence the 
resistance significantly. This is in agreement with observation reported 
in Ref. 
\onlinecite{Tsui} 
that in strong parallel fields $dR /dT \approx 0$ and conductivity of 
Si-MOSFET is temperature-independent. As a result, the symmetric part 
of resistance almost disappears. On the other hand, the  asymmetric part, 
$R_A (I_\mathrm{DC})$, is enhanced and can no longer be explained
by the influence of $V_\mathrm{DC}$. Indeed, the latter effect is
too weak and the associated term in $R_A$  should be linear in 
$I_\mathrm{DC}$ and (almost) independent of the magnetic field.

  As already mentioned in the Introduction, we suggest that the 
observed resistance asymmetry of a slot-gate 
Si-MOSFET in a parallel magnetic field should be understood in terms 
of the current-induced electron spin 
accumulation/depletion near the slot. Indeed, at $I_{\rm DC}=0$ the
magnetisation density is uniform and takes value
\begin{equation}
M_0\equiv \frac{1}{2}(n_\uparrow-n_\downarrow)=\frac{1}{2}g \mu_B \nu_0 B
\label{eq:M0}
\end{equation}
(in units of Bohr magneton $\mu_B$ per unit area). Here, $g$ is 
the gyromagnetic ratio, $n_\uparrow$($n_\downarrow$) is the spin-up (spin-down) electron density, and
\begin{equation}
\nu_0=\gamma m_*/(2 \pi \hbar^2)
\label{eq:nu0}
\end{equation}
is the electron density of states per 
spin
projection, taking into account the presence of $\gamma$ equivalent valleys
(for a Si-MOSFET, $\gamma=2$); $m_*$ is the effective mass.

\begin{figure}
\includegraphics[width=8.5cm]{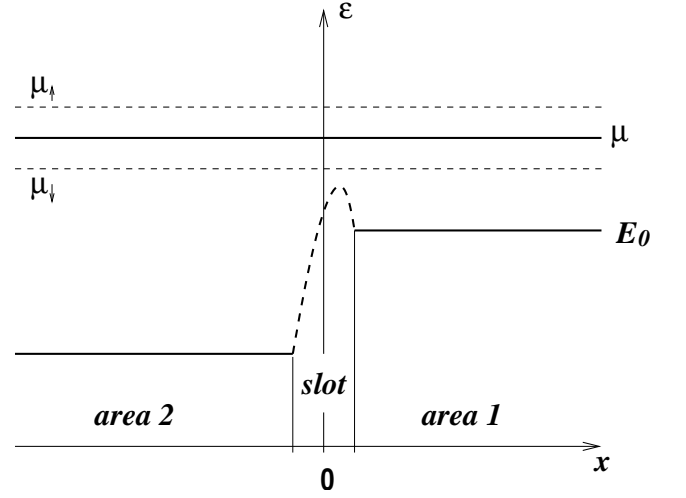}
\caption{\label{fig:energyscheme} The 2DEG in a slot-gate
MOSFET at equilibrium ($I_{\rm DC}=0$), schematic representation. In each of 
the 
two areas, the energy $E_0$
of the conduction band bottom is controlled by the corresponding gate voltage.
Chemical potential $\mu$ is uniform; applied magnetic field gives rise
to the Zeeman splitting ($\mu_{\uparrow,\downarrow}$, dashed lines).}
\end{figure}

Different electron concentrations across the slot imply  different 
degrees of spin polarisation ${\cal P}=2M_0/n$ in the presence of a magnetic field.  Within the Drude approach, one obtains
a simple relationship between electric ($j$) and spin ($s$) current 
densities away from the slot, where the state of the system remains 
uniform:     
\begin{equation}
s\equiv -\frac{e{\cal E}\tau_p}{2m_*}(n_\uparrow -n_\downarrow)=
-{\cal P}j/2e\,.
\label{eq:drudespin}
\end{equation}
Here ${\cal E}$ is the local in-plane electric field, $\tau_p$ 
is the momentum relaxation time, and $-e$ is electron charge.
Consider the case of 
$I_\mathrm{DC} > 0$, corresponding to the flow of (appropriately 
spin-polarised) electrons from area 1 with high degree of spin 
polarisation to area 2, where relative spin polarisation is smaller (Fig.
\ref{fig:energyscheme}). 
Since $j=I_{\rm DC}/d$ (where $d$ is the width of the sample) is constant,  it follows from Eq.(\ref{eq:drudespin})
that the flow of the spin density (flowing from right to left in Fig. \ref{fig:energyscheme}) is larger in area 1 than in
area 2, with the excess spin being deposited in the region around the gap.
Hence we observe that such a current causes a local increase of spin 
polarisation  near the slot, resulting in an increase of the 
overall resistance (due to the PMR effect). 
Conversely, an electron flow from area 2 to area
1 results in a spin depletion and therefore in a decreased resistance. 
While relegating a self-contained theoretical discussion to the next
section, here we quote an expression obtained in the simplest case when 
the degree
of spin polarisation is small everywhere and the spin current is continuous
at the slot. While these assumptions are at best inexact, the result
is instructive in terms of initial understanding of the data. We find  
\begin{eqnarray}
&&R_A=\frac{2 |B| I_{\rm DC}}{ed^2} \frac{n_1^{-1}-n_2^{-1}}
{\sqrt{n_1 \tau_{p,1}/
\tau_{s,1}} +\sqrt{n_2 \tau_{p,2} /\tau_{s,2}}} \times \nonumber \\ 
&&\left[\sqrt{n_1 \tau_{p,1} \tau_s,1} \frac{\partial \rho(n_1,B)}{\partial B}
+\sqrt{n_2 \tau_{p,2} \tau_s,2} \frac{\partial \rho(n_2,B)}{\partial B}
\right]+ \nonumber \\
&&+\frac{8m_*M_0^2 I_{\rm DC}}{e^3 d^2} \frac{(n_1^{-1}-n_2^{-1})^3}{(\sqrt{n_1 \tau_{p,1}/
\tau_{s,1}} +\sqrt{n_2 \tau_{p,2} /\tau_{s,2}})^2}\,
\label{eq:ra}
\end{eqnarray}
where 
$\tau_p$ is the 
momentum relaxation time, which can be roughly estimated from mobility,
Ref. \onlinecite{Shlimak12}.
The accumulated excess spin diffuses away from the slot with the rate 
controlled
by the spin relaxation time $\tau_s$ (also denoted $T_1$ in the context of 
resonance measurements). Both $\tau_p$ and $\tau_s$ depend on the carrier
density, and in Eq. (\ref{eq:ra}) we used shorthand notation, {\it viz.,}
$\tau_{p,1} \equiv \tau_p(n_1)$, etc.

The first term in Eq. (\ref{eq:ra}) describes the effect of spin accumulation or depletion on $\tau_p$ via the PMR phenomenon.
The PMR effect is parametrised by the derivatives, 
${\partial \rho(n_{1,2},B)}/{\partial B}$, which can be determined from the
data of Ref. \onlinecite{Shlimak12} using a linear fit in the carrier density. 
Studies of spin relaxation in Si/Si-Ge quantum wells 
were reported in Ref. \onlinecite{Wilamowski}, confirming that
$\tau_s$ is proportional to $\tau_p$, as expected for Dyakonov-Perel'
mechanism\cite{Dyakonov} of spin relaxation. The ratio 
$\tau_s/\tau_p$ was measured\cite{Wilamowski}  as $10^{6}$. 
Subsequent measurements yielded $\tau_s/\tau_p \sim 3\cdot 10^5$ for  
Si/Si-Ge quantum wells\cite{Tyryshkin} and $\tau_s/\tau_p \sim 2\cdot 10^5$
for a Si-MOSFET\cite{Shankar} (in the latter case, the values of carrier 
density and mobility
differed strongly from those in our measurements). 
We therefore conclude that the ratio $\tau_s/\tau_p$ is not known precisely,
leaving us with a certain freedom in the choice of the value of 
this parameter. 

In addition to these PMR-related effects, there is also another 
contribution to the resistivity, due to the
spin diffusion {\it per se}. Indeed, maintaining a non-equilibrium 
value of spin density in the region near
the slot requires a steady flow of energy to this region, resulting 
in an overall resistance increase
[cf. Eq. (\ref{eq:rdiff}) below]. The second term in Eq. (\ref{eq:ra}) is the antisymmetric part of this
additional resistance. In our range of parameter values, this term is an order of magnitude smaller than
the first one.

\begin{figure}
\includegraphics[width=9.5cm]{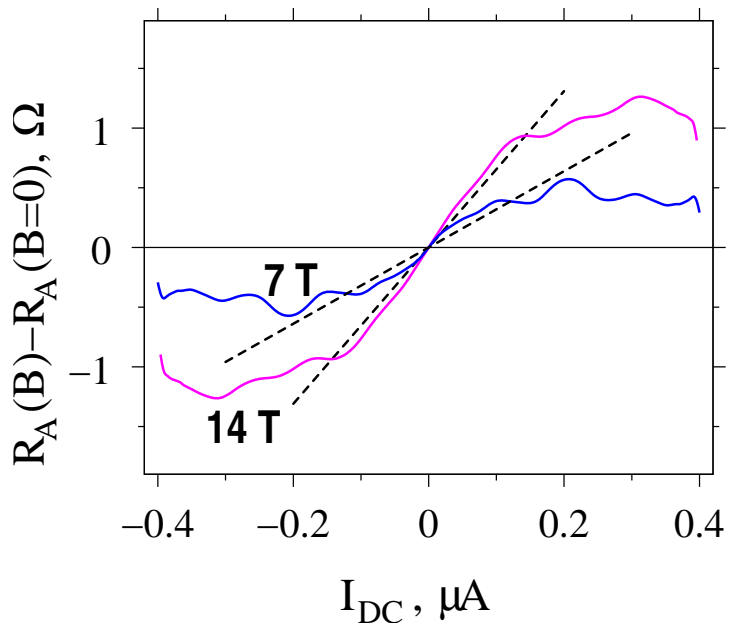}
\caption{\label{fig:linearfit} (colour online).
Antisymmetric parts of $B=7T$ and $B=14T$ data shown in Fig.
\ref{fig:datafield}, averaged over noise. 
The values of 
$R_A(I_{\rm DC})$ at $B=0$ were subtracted in order to eliminate the
contribution of the ``pinch-off'' effect. 
Dashed lines correspond to the crude theoretical result, Eq. (\ref{eq:ra}).
}
\end{figure}

We find that a perfect fit to the experimental $R_A(I_{\rm DC})$ at
small $I_{\rm DC}$ is obtained if we assume $\tau_s/\tau_p=1.7\cdot 10^{5} $ 
(see Fig. \ref{fig:linearfit}), slightly below the reported range. 

We also note the pronounced deviation of experimental curves from the
linear form of Eq. (\ref{eq:ra}) at larger $I_{\rm DC}$. This suggests
the importance of an additional, non-linear mechanism for dissipating
excess spin (of either sign) in a slot-gate MOSFET. Such a mechanism will
be introduced in Sec. \ref{subsec:boltzmann}. We will then continue with 
the  analysis of our experimental data in Sec. \ref{sec:compare}.

\section{Theoretical Model}
\label{sec:theory}

\subsection{From the Boltzmann Equation to Spin Dynamics}
\label{subsec:boltzmann}

We begin with modelling our system microscopically with the help of
a simple Boltzmann equation. Analysis of resultant macroscopic
equation for magnetisation is relegated to the next subsection.

\begin{figure}
\includegraphics[width=8.5cm]{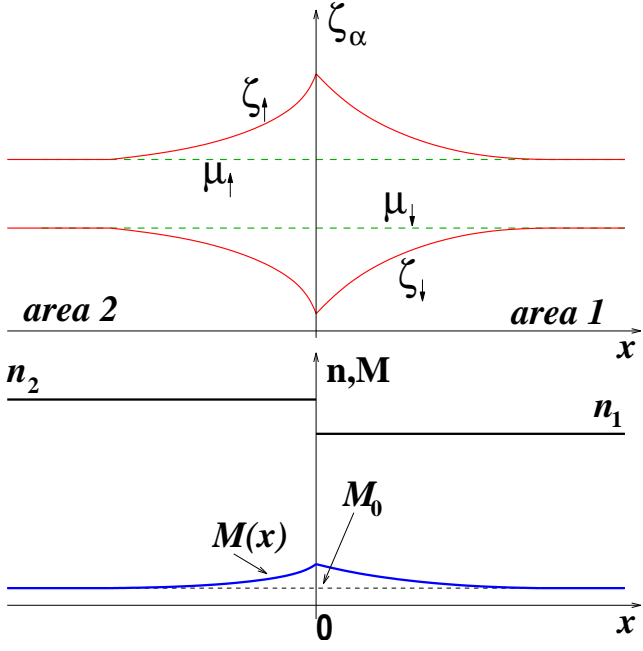}
\caption{\label{fig:zeta} (colour online). 
Schematic profiles of $\zeta_{\uparrow,\downarrow}(x)$, $n(x)$, and $M(x)$
around the slot. The width of the slot is assumed negligible, and the 
direction in which  $\zeta_{\uparrow,\downarrow}(x)$ deviate from the 
equilibrium values $\mu_{\uparrow,\downarrow}$ corresponds to $I_{\rm DC}>0$.
(see Fig. \ref{fig:deltam} below).}  
\end{figure}

We make use of the fact\cite{Wilamowski} that the spin relaxation time 
$\tau_s$ 
is much larger  than the carrier scattering time $\tau_p$. It is
this latter time which characterises the {\it momentum} relaxation of the 
system to a ``quasi-stationary'' state with the distribution function
\begin{equation}
f_\alpha(\vec{p},x)= 
\left[\exp\left(\frac{\epsilon_{\vec{p}}-\zeta_\alpha(x)}{T}\right)+
1\right]^{-1},
\,\epsilon_{\vec{p}}=\frac{p^2}{2m_*}+E_0(x)\,,
\label{eq:interim}
\end{equation}
characterised by the effective chemical potentials 
$\zeta_{\uparrow,\downarrow}(x)$.
Here, $E_0$ is the energy of the bottom of the band, which depends on the
co-ordinate $x$ (along the sample) and is determined primarily by
the gate voltage. Strictly speaking, it is also  affected by the source-drain 
bias (cf. pinch-off effect in field-effect transistors \cite{Cobbold}). 
The latter effect
gives rise to a small correction, $\Delta R_0(I_{\rm DC})$, to the measured 
resistance.
This $\Delta R_0$ and the larger term due to spin-transport effects (which
is of interest to us here) are additive. We will estimate $\Delta R_0$
phenomenologically in Sec. \ref{sec:compare} below, while presently  assuming
that the value of $E_0(x)$ is independent of the source-drain voltage.
The ``quasi-stationary'' values  of the chemical potential for the
corresponding spin species, $\zeta_{\uparrow,\downarrow}(x)$,
(which also include the Zeeman energy) are related to the local carrier 
density and magnetisation density (the latter in the units 
of $\mu_B$ per unit area) according to
\begin{equation}
n=\nu_0 (\zeta_\uparrow+ \zeta_\downarrow-2E_0)\,,\,\,\,M=\frac{1}{2}
\nu_0(\zeta_\uparrow- \zeta_\downarrow)\,
\label{eq:denmag}
\end{equation}
(see Fig. \ref{fig:zeta}), where $\nu_0$ is given by Eq. (\ref{eq:nu0}).
The quantities $\zeta_\alpha$ relax to their true equilibrium values
of $\mu_{\uparrow,\downarrow}=\mu \pm \frac{1}{2} \mu_B gB$ (here $\mu$
is the chemical potential) 
with a large 
characteristic time $\tau_s$. Below we shall see that on a smaller time scale 
(or when there is a current passing through the system), the values
of  $\zeta_\alpha$ may depend on $x$. As a consequence, magnetisation
$M(x)$ may deviate from its uniform equilibrium value $M_0$,
given by Eq. (\ref{eq:M0}).

We describe the relaxation of the system to the ``intermediate'' equilibrium, Eq.
(\ref{eq:interim}), via the Drude-type Boltzmann equation, 
\begin{eqnarray}
&&\frac{\partial \delta f_{\uparrow,\downarrow}}{\partial t} 
+\frac {\partial f_{\uparrow,\downarrow}}{\partial p_x}\left\{\frac{1}{n\tau_{dr}}{P}_{\downarrow,\uparrow}-e{\cal E}-\frac{\partial E_0}{\partial x}\right\}+ 
\frac {\partial f_{\uparrow,\downarrow}}{\partial\zeta_{\uparrow,\downarrow}}
\times \nonumber \\
&&\times \left\{\frac{\partial\zeta_{\uparrow,\downarrow}}{\partial x} -    
\frac{\partial E_0}{\partial x} \right\}\frac{p_x}{m_*} = - \frac{\delta f_{\uparrow,\downarrow}}{\tau_p}- \frac {n_{\downarrow,\uparrow}
\delta f_{\uparrow,\downarrow} }{n \tau_{dr}}\,.
\label{eq:kinetic}
\end{eqnarray}
Here, $\delta f_\alpha (t, x, \vec{p})$ is the non-equilibrium part of the 
distribution function, $\tau_p$ is the momentum relaxation time (assumed
to be spin-independent), 
and $\tau_{dr}$  the spin-drag time. The terms 
containing ${\partial E_0}/{\partial x}$ cancel,
corresponding to zero current in the absence of the source-drain electric
field ${\cal E}$. The
quantities $n_\alpha$ and $P_\alpha$ are the electron density and the net 2DEG
momentum density for the
corresponding spin species,
\begin{equation}
n_\alpha = (\zeta_\alpha-E_0) \nu_0 \,,\,\,\,\, P_\alpha= \gamma \int p_x \delta f _\alpha \frac{ d^2 p}{(2 \pi \hbar)^2} \,.
\end{equation}
The latter is related to the two-dimensional charge and spin current densities 
via
\begin{equation}
j= - \frac{e}{m_*} (P_\uparrow+P_\downarrow)\,,\,\,\,
s=\frac{1}{2m_*}(P_\uparrow-P_\downarrow)  \,,
\end{equation}
where spin is again measured in units of Bohr magneton.
The Coulomb spin-drag effect\cite{DAmico_drag,Flensberg} gives rise to a ``drag'' force appearing
on  the l. h. s. of Eq. (\ref{eq:kinetic}), and to another channel of momentum
relaxation corresponding to the second term on the r. h. s.. 

In the steady state at low $T$, multiplying Eq. (\ref{eq:kinetic}) by $p_x$
and integrating yields

\begin{equation}
\left(e{\cal E}+\frac{\partial \zeta_{\uparrow,\downarrow}}{\partial x}-
\frac{P_{\downarrow,\uparrow}}{n \tau_{dr}}\right) n_{\uparrow,\downarrow} =
-\frac{1}{\tau_p}P_{\uparrow,\downarrow}-\frac{n_{\downarrow,\uparrow}}
{n \tau_{dr}}P_{\uparrow,\downarrow} \,.
\end{equation}
Using $n_{\uparrow,\downarrow}=\frac{1}{2}n \pm M$, we next find

\begin{eqnarray}
&&\left( e{\cal E}+\frac{\partial E_0}{\partial x}
 +\frac{1}{2 \nu_0} \frac {\partial n}{\partial x}\pm \frac {1}{\nu_0}
\frac{\partial M}{\partial x}\right) \left(\frac{1}{2}n \pm M \right)
=
\nonumber \\
&&=\frac{m_*}{\tau_p}(\frac{1}{2e}j\mp s) \mp \frac{m_*}{n \tau_{dr}}(ns+\frac{1}{e}Mj)\,. 
\label{eq:macro}
\end{eqnarray} 
Summing the two equations (\ref{eq:macro}) yields an expression
\begin{equation}
e{\cal E} = \frac{m_*}{ne\tau_p} j- \frac{1}{2 \nu_0}  \frac{\partial n}{\partial x}- \frac{2}{n\nu_0}M \frac{\partial M}{\partial x}+
\frac{\partial E_0}{\partial x} 
\label{eq:electric}
\end{equation}
for the electric field ${\cal E}$. In principle, this should be solved 
together with
the appropriate Poisson equation and with the spin dynamic equations, to 
determine $n(x)$, $M(x)$, and ${\cal E}(x)$ self-consistently. 

We note that our experimental setup is reminiscent of the one previously
considered in the literature\cite{Privman,Villegas}, whereby
the doping level in a (three-dimensional) semiconducting sample is varied
abruptly as a function of $x$, resulting in the carrier density jump at $x=0$.
Electrical current is passed along the $x$ axis,
and while there is no external magnetic field, spin current is injected into
the semiconductor from a ferromagnetic tip located away from the 
$x=0$ plane. 
It was suggested\cite{Privman,Villegas,Csontos} that taking into account
the non-trivial $x$ dependence of the electric field and of the carrier 
density (as 
dictated by the Poisson equation) is essential for correctly describing 
the system. Here, we wish to argue that the latter complication does not arise
in the present case.

When supplemented with the Poisson equation, Eq. (\ref{eq:electric}) 
leads to a drift-diffusion
equation for the electric field ${\cal E}$. Our current densities 
are well within the diffusive regime of this
equation.  Furthermore, we find that in our case, the associated diffusion length
is of the order of a few nm, which is much smaller than any relevant length scale. The "smearing"
of $n(x)$ is therefore insignificant. 
We  conclude that in the range of 
parameter values of interest to us one can neglect the feedback effect of 
$\partial {\cal E}/\partial x$ on $n$ and $M$ and omit the Poisson equation 
altogether. This amounts to 
assuming 
\begin{equation}
n(x)=\left\{ \begin{array}{ll} n_1\,,& x>0\,,\\
n_2\,,& x<0\,,
\end{array}\right.
\label{eq:n}
\end{equation}
where $n_1$ and $n_2$ are 
the 2DEG densities as set by selecting the appropriate gate voltages at $j=0$ 
and $B=0$. 
The width of the slot, 90 nm, is much larger than the Fermi wavelength
$\hbar/p_F$ for our values of $n_{1,2}$, hence the changes in $E_0$ and
$n$ affect the quantum-mechanical carrier motion only adiabatically. 
Accordingly, one can assume that the carriers pass across the slot region in 
a ballistic fashion [as opposed to tunnelling; treating the slot as a 
tunnelling barrier with a finite spin-dependent conductance yields only a 
quantitative change in the resultant $R(I_{\rm DC})$ dependence]. Note also 
that we do 
not attempt to model the profile of
$n(x)$ [and $E_0(x)$] within the slot, since the slot width is smaller than the
characteristic length scale of the spin dynamics (spin diffusion length).  
The $\partial E_0/\partial
x$ term in Eq. (\ref{eq:electric}) is compensated over short distance by
the density variation, $\partial n/\partial
x$, (as described by the Poisson equation and independently of the 
source-drain bias) and both terms can be
dropped.  Eq. (\ref{eq:electric}) then
merely yields the value of ${\cal E}$ as a function of current density $j$ 
(a constant playing the role of  experimental control parameter) and 
magnetisation $M(x)$. The latter is determined by the spin dynamics, to 
which we will turn now.

Subtracting the two equations (\ref{eq:macro}) from each other, we find the
following expression for the spin current [in units of $\mu_B$:
\begin{equation}
s=-\frac{Mj}{en} - \frac{\tau}{2m_* n \nu_0} (n^2-4M^2)
\frac{\partial M}{\partial x}\,,
\label{eq:spincurrent}
\end{equation}
where 
\begin{equation}
\frac{1}{\tau}=\frac{1}{\tau_p}+\frac{1}{\tau_{dr}}\,.
\end{equation}
Thus, the effect of the Coulomb spin drag on spin dynamics in our case
consists in a mere relaxation time renormalisation\cite{Flensberg}.
The precise value of $1/\tau_{dr}$ is not known, but is 
expected\cite{DAmico_drag}
to be small at low temperature $T$. Therefore when comparing our theoretical results with
the experimental data in Sec. \ref{sec:compare},
we will assume $\tau \approx \tau_p$. We also note
that the last (diffusive) term in Eq. (\ref{eq:spincurrent}) vanishes in 
a uniform system ($\partial M/\partial x=0$) or for the case of complete 
spin polarisation ($n=2M$).

The continuity equation for  magnetisation reads
\begin{equation}
\frac{\partial M}{\partial t} = - \frac{\partial s}{\partial x}- 
\frac{M-M_0}{\tau_s}\,.
\label{eq:spincont}
\end{equation} 
The spin relaxation time, $\tau_s$, is due primarily to the Dyakonov-Perel'
mechanism\cite{Wilamowski,Dyakonov}. It does depend on $n$, but an increase 
of temperature (which might occur due to Joule heating) does not affect
the value of $\tau_s$ as long as $T$ is small compared to the Fermi 
energy\cite{Dyakonov}. 
Likewise, the effect of the
electrical current on $\tau_s$ is negligible if the carrier
drift velocity is much smaller than the Fermi velocity. Since the latter two 
conditions are certainly met in our experiments, we can assume that $\tau_s$ is
determined solely by the carrier density $n$.

In order to proceed with solving Eqs. (\ref{eq:spincurrent}) and 
(\ref{eq:spincont}) in the steady-state (see the next subsection), we need 
to specify the boundary condition for $M$ at the point of density jump, $x=0$.  
This can be done by replacing the step in Eq. (\ref{eq:n}) with a smooth
density change from $n_2$ to $n_1$, occurring in the range $|x|<x_0$, and
taking the limit $x_0 \rightarrow 0$. Eq. (\ref{eq:spincurrent}) is valid
for smooth $n(x)$ and $M(x)$, and must yield a finite value of spin current
$s$. Since it includes $\partial M/\partial x$, but not $\partial n/\partial
x$ [which is divergent at $x_0 \rightarrow 0$ and might have compensated
for a jump in $M(x)$ in this limit], we conclude that 
{\it magnetisation $M(x)$ must be continuous at $x=0$} (cf. Ref. \onlinecite{Pershin}). One can arrive at the
same conclusion by noticing that the two chemical potentials $\zeta_{\uparrow,\downarrow}$
must be continuous at the slot. This would be modified when  a finite 
tunnelling conductance through the slot is assumed, resulting in a 
current-dependent magnetisation step. As noted above, this modification
does not affect our results in a qualitative way, hence infinite slot
conductance will be assumed forthwith. 

We further emphasise that  $M(x)$ must be continuous at $x=0$ only as long as
the spin-polarisation on both sides of the slot remains incomplete, $M^2<
n^2/4$. This is due to the fact that the second term in 
Eq. (\ref{eq:spincurrent}) vanishes at $|M|=n/2$. Whenever full polarisation 
is attained on either side of the slot, the magnetisation can suffer a jump 
at $x=0$, and the limiting
value of $M$ on the opposite side is determined by the boundary condition for
spin current at the slot (see below). Presently, however, we shall be 
interested in the case
of incomplete polarisation only.

Our data imply that in addition to the Dyakonov-Perel' mechanism
[which is linear, see Eq. (\ref{eq:spincont})], another route of spin
dissipation is involved,  accounting for the saturation of the
antisymmetric resistance $R_A$ at larger $I_{\rm DC}$, as seen in
Figs. \ref{fig:datafield} and \ref{fig:linearfit}. This additional 
mechanism must 
be non-linear. While in
principle it could arise from a sublinear correction to the transport
and spin relaxation equations considered above, this would require
much higher values of $I_{\rm DC}$. We conclude that  a strongly
non-linear ``leakage'' of spin near the slot into a ``spin reservoir''
must be present. Assuming that the current does not flow through
this reservoir, the chemical potentials for spin-up and spin-down 
electrons within the reservoir would retain their equilibrium values, 
$\mu_{\uparrow,\downarrow}$. There are several possible realisations
of this mechanism, and in the remaining part of this subsection we will
describe two examples.
 
\begin{figure}
\includegraphics[width=8.5cm]{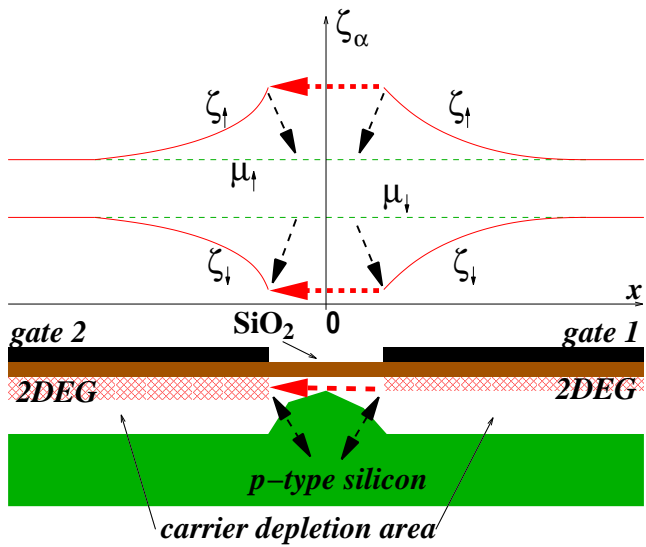}
\caption{\label{fig:tunnel} (colour online). 
Electron tunnelling processes at the slot. 
Direct tunnelling between the two areas of 2DEG is shown by thick horizontal
arrows. In addition, we include tunnelling between the
2DEG and p-type Si (slanted arrows), which arises due to a difference
between the quasi-stationary electrochemical potentials 
$\zeta_{\uparrow,\downarrow}$ for the spin-up and spin-down electrons in
2DEG and the corresponding values $\mu_{\uparrow,\downarrow}$ in the bulk.
This difference, in turn, is due to a non-zero current $I_{\rm DC}$ through the
2DEG; schematic profiles of $\zeta_{\uparrow,\downarrow}(x)$ in the
figure correspond to $M(x)>M_0$ near the slot, such as for $n_1<n_2$ and
$I_{\rm DC}>0$ (see Fig. \ref{fig:deltam} below).}  
\end{figure}

We notice that
in the vicinity of the slot, due to the absence of the gate potential,
the region of $p$-type Si approaches the surface of the sample,
thus potentially enabling electron tunnelling between the 2DEG and the 
bulk. In the absence of $I_{\rm DC}$, carriers in the bulk are at
equilibrium with the 2DEG, which means that the chemical potentials
for spin-up and down electrons have the same respective values, 
$\mu_{\uparrow,\downarrow}$.
At $I_{\rm DC}\neq 0$, these bulk values of chemical potentials do not
change, whereas those of 2DEG acquire the respective quasi-stationary values,
$\zeta_{\uparrow,\downarrow}(x)$. Since the carrier density in 2DEG
does not change, one finds 
$\zeta_{\uparrow}(x)-\mu_\uparrow=\mu_\downarrow - \zeta_{\downarrow}(x)$. 
For the tunnelling current density $j^{(t)}_\alpha$ of
electrons with spin $\alpha=\uparrow,\downarrow$ from the 2DEG into the 
bulk we write
\begin{equation}
j^{(t)}_\alpha= G_\alpha \cdot [\mu_\alpha-\zeta_\alpha(0)]\,,
\label{eq:tuncur}
\end{equation}
where the conductance
\begin{equation}
G_\alpha=G_0+K\cdot [\mu_\alpha-\zeta_\alpha(0)]^2 
\label{eq:tuncon}
\end{equation}
is assumed to have a spin-independent value $G_0$ in the Ohmic limit. The
corresponding tunnelling processes are shown schematically in Fig.
\ref{fig:tunnel}. 
In writing Eqs. (\ref{eq:tuncur}--\ref{eq:tuncon}), we
make use of the continuity of $\zeta_{\uparrow,\downarrow}(x)$ at the slot, 
$\zeta_{\uparrow,\downarrow}(-0)=\zeta_{\uparrow,\downarrow}(+0)$. We
denote the corresponding limiting value $\zeta_{\uparrow,\downarrow}(0)$,
as the width of the slot is negligible from the viewpoint of macroscopic
equations analysed below. We see
that $j^{(t)}_\uparrow$ and $j^{(t)}_\downarrow$ cancel each other, yet 
there arises a spin current from the 2DEG into the bulk, with the density
\begin{equation}
s^{(t)}=\Gamma \cdot[M(0)-M_0]\,,\,\,\,\,\,
\Gamma=\frac{G_0}{e\nu_0}+\frac{K}{e\nu_0^3}
[M(0)-M_0]^2
\label{eq:tunspin}
\end{equation} 
[cf. Eq. (\ref{eq:denmag})]. Thus, the boundary condition for spin 
current at the slot
takes form 
\begin{equation}
s(+0)=s(-0)-s^{(t)}\,,
\end{equation}
where the limiting values $s(\pm 0)$ of  the 2DEG spin current density to the
right and to the left of the slot are given by Eq. (\ref{eq:spincurrent}).

\begin{figure}
\includegraphics[width=8.9cm]{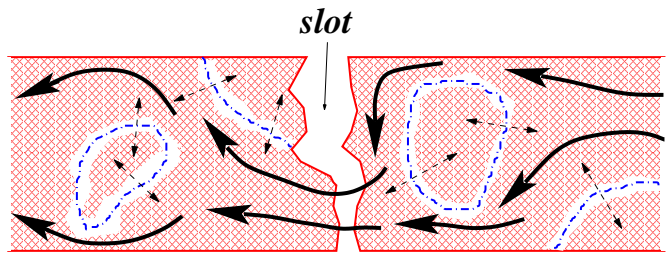}
\caption{\label{fig:puddles}(colour online).
Inhomogeneities in the 2DEG (schematic view from above): hatched areas 
correspond to the 2DEG; dashed-dotted lines, to potential barriers. Solid lines
show the electron flow corresponding to the source-drain current (``stream''), 
dashed double-arrowed lines -- tunnelling to/from
isolated areas of the 2DEG (``puddles'').}  
\end{figure}

We note that the physics associated with Eqs. 
(\ref{eq:tuncur}--\ref{eq:tunspin})
is not restricted to the specific case of tunnelling into the p-type Si, as
described above. Another alternative possibility is related to the fact that
the current flow {\it within} 2DEG (and especially near the slot)  is not 
necessarily uniform; instead, there
might exist sizable regions of 2DEG which do not participate in 
conduction; these ``puddles'' would be separated from the ``stream'', where the
current is flowing, by relatively low tunnelling barriers (see Fig. 
\ref{fig:puddles}). Due to the absence 
of current, the respective chemical potentials for
the two spin species in the ``puddles'' retain their unperturbed values, 
$\mu_\alpha$. Hence the ``puddles'' would play exactly
the same role of ``spin reservoir'' as the p-type Si in the previous scenario.
What is essential for us here is that there is a ``leakage'' of spin current
from the source-drain current flow into the reservoir, and that this leakage 
depends on $\mu_\alpha - \zeta_{\alpha}(x)$ in a strongly non-linear fashion
[cf. Eqs. (\ref{eq:tuncon}--\ref{eq:tunspin})]. The reservoir must be 
sufficiently large to
allow for efficient relaxation of the incoming excess spin.

\subsection{Spin Drift-Diffusion Phenomena, and the Effect on the Sample
Resistance}
\label{subsec:drift}

Our first objective here is to find the steady-state profile of magnetisation
$M$ as a function of the co-ordinate $x$ for given values of current $j$ and
carrier densities $n_{1,2}$. 
Combining Eqs. (\ref{eq:spincurrent}) and 
(\ref{eq:spincont})  yields the spin drift-diffusion equation,
\begin{equation}   
\alpha \frac{\partial}{\partial x} \left[ \left( \frac{1}{4}n^2-M^2 \right) \frac{\partial M}{\partial x} \right]
+\frac{1}{2} j \frac{\partial M}{\partial x}- \beta (M-M_0)=0,
\label{eq:spindd}
\end{equation}
where
\begin{equation}
\alpha=\frac{e \tau}{m_* \nu_0}
\,,\,\,\,\,\,\,\,\,\,\,
\beta=\frac{en}{2\tau_s}\,.
\end{equation}

We note that spin drift-diffusion equations were derived earlier by 
Yu and Flatt\'{e} for non-degenerate semiconductors\cite{Flatte}
and by D'Amico for the degenerate case\cite{DAmico_drift}; an interface
problem similar to the present one was considered in Ref. \onlinecite{Pershin}.
In all these cases, the linearisation in $\delta M =M(x)-M_0$ was performed,
resulting in a linear drift-diffusion equation. On the other hand, the
first term of our Eq. (\ref{eq:spindd}) is explicitly non-linear. In
addition to enforcing a physical constraint, $|M(x)|<n/2$, this
non-linearity affects the subleading (in $j$) terms even for small
$|\delta M| \ll n$, which will be important for us here.

At $j=0$, Eq.  (\ref{eq:spindd}) is solved by  $\delta M \equiv 0$; other
solutions may exist, but these appear irrelevant for the case at hand. As 
mentioned
above, within the present macroscopic description 
the carrier density $n$ [see Eq. (\ref{eq:n})], as well as $\tau$ and $\tau_s$ 
which 
depend
on $n$, suffer a jump at $x=0$. In the following, the subscript 1 (2) refers 
to the
quantities characterising the $x>0$ ($x<0$) part of the sample.

We are interested in the {\it diffusive} regime of small current 
densities, $|j| \ll j_{{\rm cr},i}$
for $i=1,2$, where
\begin{equation}
(j_{{\rm cr},i})^2=4 \alpha_i \beta_i
(n_i^2-4M_0^2)=\frac{2e^2n_i \tau_i}{m_* \nu_0 \tau_{{\rm s}\,i}} 
(n_i^2-4M_0^2)\,.
\label{eq:jcr}
\end{equation}
We estimate that for our system, $j=j_{{\rm cr},i}$ would correspond to 
a net current $I_{\rm DC}$
which is an order of magnitude larger than our operational values\cite{jcr}. 

In the diffusive regime, 
$\delta M(x)$ is small everywhere, and Eq. (\ref{eq:spindd}) is easily solved
by iterations. Keeping terms of up to  second order in $j$, one finds
the appropriate solution, decaying exponentially at
large $|x|$,
\begin{equation}
\delta M= \left\{ \begin{array}{ll} C_1 \exp(\lambda_1 x)+A_1
 \exp(2\lambda_1 x)\,,& x>0\,,\\ \, & \, \\
C_2 \exp(\lambda_2 x)+A_2
 \exp(2\lambda_2 x)\,,& x<0\,,
\end{array}\right.
\label{eq:deltam}
\end{equation}
where
\begin{equation}
\lambda_{i}= \mp 2 \sqrt{\frac{\beta_i}{\alpha_i(n_i^2-4M_0^2)}}\left(1 \pm
\frac{j}{j_{{\rm cr},i}} \right)\,
\label{eq:lambda}
\end{equation}
and
\begin{equation}
A_i \approx \frac{16}{3}M_0 C_i^2 \frac{1}{n_i^2-4M_0^2}\,.
\label{eq:Ai}
\end{equation}

Expressions for $C_i$ are found from the boundary conditions for spin current
and magnetisation, as discussed in the previous subsection.
To leading order in $j/j_{{\rm cr},i}$, we find $C_1=C_2=C_0$ where
the quantity $C_0$ is the solution of 
\begin{equation}
C_0=2M_0j \left(\frac{1}{n_1}-\frac{1}{n_2}\right)
\left(\frac{j_{{\rm cr},1}}{n_1}+\frac{j_{{\rm cr},2}}{n_2}
+\frac{2 G_0}{\nu_0}+\frac{2 K}{\nu_0^3}C_0^2\right)
^{-1}\,.
\label{eq:c0}
\end{equation}
To the required accuracy, we can substitute $C_i \rightarrow C_0$ in Eq.
(\ref{eq:Ai}).
The subleading terms in $C_i$ are given by
\begin{widetext}
\begin{equation}
C_i-C_0= 
C_0\left[\frac{8C_0 M_0 j_{{\rm cr},1}}{3 n_1(n_1^2-4M_0^2)}+
\frac{8 C_0 M_0j_{{\rm cr},2}}
{3n_2(n_2^2-4M_0^2)}+ j\left(\frac{1}{n_1}-\frac{1}{n_2} \right) \right]
\left[\frac{j_{{\rm cr},1}}{n_1}+\frac{j_{{\rm cr},2}}{n_2}+
\frac{2G_0}{\nu_0}+\frac{6 K}{\nu_0^3}C_0^2\right]^{-1} -A_i\,.
\label{eq:c1}
\end{equation}

Next, we must use Eq. (\ref{eq:electric}) to express the potential difference 
between 
the voltage contacts (located at $x=-L_2$
and $x=L_1$ with $L_{1,2}\gg |\lambda_{1,2}|^{-1}$) as
\begin{equation}
V \equiv \int_{-L_2}^0 {\cal E}(x) dx + \int^{L_1}_0 {\cal E}(x) dx= 
j\left\{ \int_{-L_2}^0 \rho\left[n_2,M(x)\right] dx + 
\int^{L_1}_0  \rho\left[n_1,M(x)\right]dx  \right\}+
\frac{1}{e\nu_0}\left(\frac{1}{n_1}-
\frac{1}{n_2} \right)\left\{ [M(x=0)]^2-M_0^2 \right\}\,.
\label{eq:voltage}
\end{equation} 
\end{widetext}
Here, the first term on the r.\ h.\ s. corresponds to the Ohm's law,
and in writing it we take into account the fact that the well-known positive 
magnetoresistance of the 2DEG in a parallel magnetic field B is of spin
origin. In other words, 
$\rho$ depends on $B$  via  
the field dependence of magnetisation $M$, {\it viz.,} $\rho=\rho[n,M(B)]$,
or $\partial \rho/\partial M=2(\nu_0 \mu_B g)^{-1}\,\partial \rho(n)/\partial B$.
The last term in Eq. (\ref{eq:voltage}) originates from the third term 
in Eq. (\ref{eq:electric}); essentially, this is the additional voltage
required to maintain the (non-equilibrium) non-uniform profile of
chemical potentials $\zeta_{\uparrow,\downarrow}(x)$ which results
in a non-zero $\delta M(x)$ [see Eq. (\ref{eq:denmag})].

\begin{figure}
\includegraphics[width=8cm]{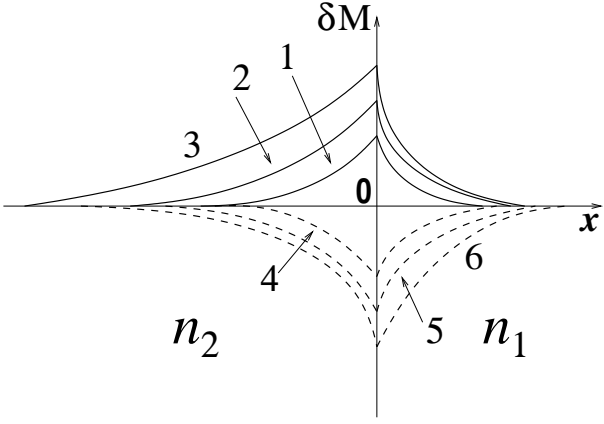}
\caption{\label{fig:deltam}  
Schematic behaviour of $\delta M(x)$ for different values of $I_{\rm DC}$,
assuming $n_2>n_1$. Curves 1 to 3 correspond to increasing values of
$I_{\rm DC}>0$, whereas curves 4 to 6 illustrate the effect of
increasing  $|I_{\rm DC}|$ for $I_{\rm DC}<0$. Slight difference in the shape of the curves for different
values of $I_{\rm DC}$ reflects the presence of a
small current-dependent correction in Eq. (\ref{eq:lambda}).}  
\end{figure}

Dynamic resistance can be found as a derivative $R=dV/dI_{\rm DC}$ of the
voltage, Eq. (\ref{eq:voltage}), with respect to the net dc-current 
$I_{\rm DC}=jd$ (where $d$ is the width of the sample). We first consider
the case of very small current densities, $j \ll j_{\rm sat}$, where
\begin{equation}
8 K M_0^2j_{\rm sat}^2 \sim \nu_0^3 \left(\frac{j_{{\rm cr},1}}{n_1}+
\frac{j_{{\rm cr},2}}{n_2}+
\frac{2G_0}{\nu_0} \right)^3 \left(\frac{1}{n_1}-\frac{1}{n_2} \right)^{-2}  
\,.
\label{eq:jsat}
\end{equation}
We will have to assume that the quantity $K$ [parametrising the non-linearity
of  tunnelling into the p-type Si, see Eq.(\ref{eq:tuncon})] is large, so 
that this 
condition is more restrictive than $j \ll j_{{\rm cr},i}$, 
cf. Eq. (\ref{eq:jcr}). Nevertheless, with $j \ll j_{\rm sat}$ one can 
neglect the
$K$ terms on the r.\ h.\ s. of Eqs. (\ref{eq:c0}) and (\ref{eq:c1}), 
enabling analytical calculation. The dependence of $\delta M(x)$ on
$I_{\rm DC}$ in this regime is shown schematically in Fig.  \ref{fig:deltam}.

Substituting Eq. (\ref{eq:deltam}) into (\ref{eq:voltage}) and expanding
to quadratic terms in the net dc-current $I_{\rm DC}=jd$, 
we obtain 
\begin{equation}
V=  I_{\rm DC} R(0) +\frac{1}{2} I_{\rm DC}^2
\frac{\partial R}{ \partial I_{\rm DC}}\,,
\label{eq:voltageexp}
\end{equation}
where the differential resistance at zero current is
\begin{widetext}
\begin{equation}
R(0)=\frac{1}{d}\left[ \rho(n_1,M_0)L_1+ \rho(n_2,M_0)L_2 \right]
+\frac{4M_0^2}{e \nu_0 d} \left(\frac{1}{n_1}-\frac{1}{n_2}\right)^2
\left(\frac{j_{{\rm cr},1}}{n_1}+\frac{j_{{\rm cr},2}}{n_2}
+ \frac{2G_0}{\nu_0}\right)^{-1}\,.
\label{eq:rdiff}
\end{equation}
We see that the resistance of two 2DEG areas connected in
series is increased due to spin effects, as implied by the presence of
the second term. It originates from the third (spin-diffusion) term in
Eq. (\ref{eq:electric}), and  corresponds to a linear increase
of $|\delta M(x=0)|$ in the diffusive regime (cf. Fig.  \ref{fig:deltam}).
This non-equilibrium distribution of $M(x)$ is maintained by the
current flow, which increases the net resistance.
In addition, a deviation from the Ohm's law is obtained,
{\it viz.}, $R(I_{\rm DC})=R(0)+ I_{\rm DC}\partial R/\partial I_{\rm DC}$, where
\begin{eqnarray}
&&\frac{\partial R}{\partial I_{\rm DC}} = \frac{M_0}{ d^2} 
\left(\frac{1}{n_1}-\frac{1}{n_2}\right)
\left(\frac{j_{{\rm cr},1}}{n_1}+\frac{j_{{\rm cr},2}}{n_2}
+ \frac{2G_0}{\nu_0}\right)^{-1}\,
\left(\frac{j_{{\rm cr},1}}{\beta_1}\frac{\partial \rho(n_1,M_0)}{\partial M}+
\frac{j_{{\rm cr},2}}{\beta_2}\frac{\partial \rho(n_2,M_0)}{\partial M}
\right)+ 
\nonumber \\
&&+
\frac{16M_0^2}{3 \nu_0 e d^2} \left(\frac{1}{n_1}-\frac{1}{n_2}\right)^3
\left(\frac{j_{{\rm cr},1}}{n_1}+\frac{j_{{\rm cr},2}}{n_2} +
\frac{2G_0}{\nu_0} \right)^{-3}
\left(\frac{3n_1^2-4M_0^2}{n_1^2 -4 M_0^2}\frac{j_{{\rm cr},1}}{n_1}+
\frac{3n_2^2-4M_0^2}{n_2^2 -4 M_0^2}\frac{j_{{\rm cr},2}}{n_2}
+ \frac{6G_0}{\nu_0}\right)\,.
\label{eq:deltardiff}
\end{eqnarray}   
\end{widetext}
Here, the first term is due to the positive magnetoresistance
of the 2DEG (caused by the magnetisation change, $M_0 \rightarrow
M_0+\delta M(x)$, as explained above); the second term is
the sublinear contribution of the third term in
Eq. (\ref{eq:electric}). In the
appropriate limit of $M_0 \ll n_{1,2}$ and $G_0 \rightarrow 0$,
Eq. (\ref{eq:electric}) yields Eq. (\ref{eq:ra}). For our range
 of parameter values the first term (which is roughly linear
in $M_0$ and hence in $B$) dominates. This agrees with the experimental
$R_A(B)$ at small $I_{DC}$, as shown in Fig. \ref{fig:datafield}.

We see that the resistance indeed acquires an asymmetric contribution, as
seen in Fig. \ref{fig:datafield}. 
When the net current is
small, this contribution is linear in $I_{\rm DC}$, as per Eq. 
(\ref{eq:deltardiff}). When $|I_{\rm DC}|$  becomes
comparable to $j_{\rm sat}d$ [cf. Eq. (\ref{eq:jsat})], the increase
of $|\delta M(x)|$ with $|I_{\rm DC}|$ slows down and becomes sublinear.
This is the origin of saturation in the asymmetric contribution
to resistance as seen in the experimental data, Fig. \ref{fig:datafield}.
The value of the resistance $R$ in this region can be calculated as
\begin{equation}
R(I_{\rm DC})= \partial V(I_{\rm DC})/\partial I_{\rm DC}\,,
\label{eq:rgen}
\end{equation}
where $V(I_{\rm DC})$ is given by Eq. (\ref{eq:voltage}), and its evaluation involves numerically
solving Eq. (\ref{eq:c0}) for $C_0$.
Typical profiles of the resultant $R_A(I_{\rm DC})$ will be shown in
the next section, where these will be compared against the experimental
results.

The sublinear behaviour of $R(I_{\rm DC})$ is due to the non-linear tunnelling to the "spin reservoir",
 Eq. (\ref{eq:tuncur}--\ref{eq:tuncon}), which results in a slower growth of the "effective spin injection" rate  with current 
at larger $|I_{\rm DC}|$. One can readily see this analytically in the limiting case of $j_{\rm sat} \ll |j| \ll j_{{\rm cr},i}$, when
the last term in the denominator of Eq. (\ref{eq:c0}) dominates, leading to $C_0 \propto j^{1/3}$. In this regime, we find
that to leading order, the antisymmetric part of the resistivity, $R_A(I_{\rm DC})$, is proportional to  $I_{\rm DC}^{1/3}$.

\section{Comparison with the Experimental Data}
\label{sec:compare}

Here, we attempt a detailed comparison of our experimental data 
(Sec. \ref{sec:overview}) with the theory developed in Sec. \ref{sec:theory}.
Our focus will be on the {\it antisymmetric} part of both theoretical and
experimental results. This is because  the symmetric part can be affected by
additional physical mechanisms,
which are unrelated to spin transport and are therefore of no interest to us
here. These include the non-linearity in the slot transmission coefficient,
and the Joule heating (although, as mentioned above, the effect of heating on resistivity is strongly suppressed when
a magnetic field is applied). The antisymmetric part, on the contrary, is due
mostly to the spin transport processes as discussed theoretically in Sec. 
\ref{sec:theory}, with a smaller antisymmetric contribution due to the effect
of the source-drain potential on the 2DEG carrier density (pinch-off). 

\begin{figure}
\includegraphics[width=9.0cm]{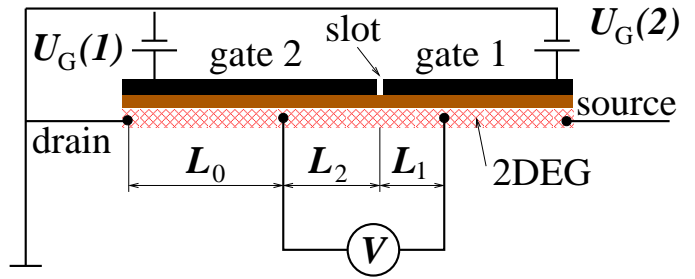}
\caption{\label{fig:pinchoff} (colour online).
Electrical connexions of the sample, with $L_0= 250$ $\mu$m, $L_2=150$ $\mu$m, and $L_1=30 $ $\mu$m. At $I_{\rm DC}=0$,
the 2DEG carrier density under gate 1 (gate 2) equals $n_1$ ($n_2$). When a source-drain bias is applied, these density values
vary slightly, resulting in a change of the corresponding resistivities.
}
\end{figure}

The latter contribution, present also at $B=0$, can be evaluated based on the
electrical connexion scheme, shown in Fig. \ref{fig:pinchoff}. 
At $I_{\rm DC}=0$,  the carrier densities in 
the 2DEG are determined by the respective gate voltages, 
yielding the values of resistivity in the two parts
of the sample $\rho_{1,2}=\rho[U_G(1,2)]$. At $I_{\rm DC} \neq 0$, 
the resultant electrical potential $\phi(x)$
within the 2DEG layer
is added to the gate voltage, and the resistivity acquires a weak dependence on the coordinate $x$, {\it viz.},
$\tilde{\rho}_{1,2}(x)=\rho[U_G(1,2)+\phi(x)] \approx \rho_{1,2}+ \phi(x) \partial \rho_{1,2}/\partial U_G$. The overall resistance change between the voltage contacts is linear 
in $I_{\rm DC}$,
\begin{eqnarray}
\Delta R_{\rm po}&=&-\frac{I_{\rm DC}}{d^2} \left\{
(L_2^2+2L_2 L_0) \rho_2 \frac{\partial \rho_2}{\partial U_G} +
\right. \nonumber \\ 
&&+\left. \left[ 2(L_0+L_2)L_1 \rho_2 +L_1^2 \rho_1 \right]
\frac{\partial \rho_1}{\partial U_G} \right\}\,,
\label{eq:pinchoffcorr}
\end{eqnarray}
and must be added to Eqs. (\ref{eq:deltardiff}--\ref{eq:rgen}) when comparing the latter with the experimental data. We note that in the
$B=0$ case Eq. (\ref{eq:pinchoffcorr}) accounts for the entire antisymmetric part of the resistance, and indeed a rather
accurate fit to the $B=0$ data in Fig. \ref{fig:datafield} c is obtained.

\begin{figure*}
\includegraphics[width=18.0cm]{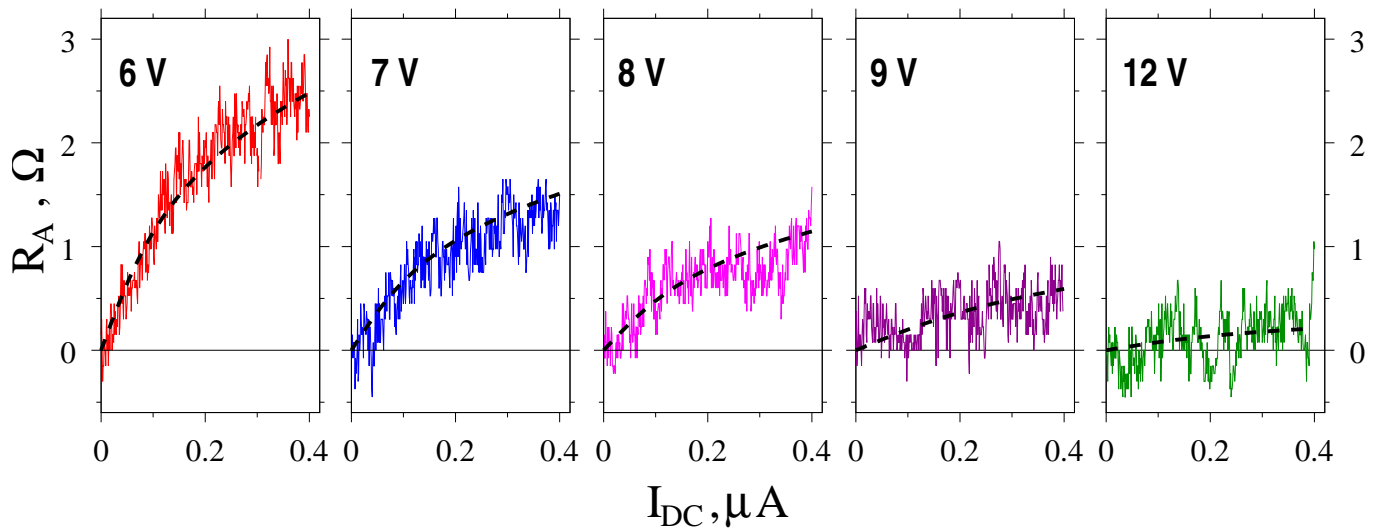}
\caption{\label{fig:gatefit} (colour online).
Antisymmetric parts of the data obtained for different values of $U_G(1)$, as
shown in Fig. \ref{fig:datagate}. The dashed lines correspond to the
theoretical result,  Eqs. (\ref{eq:deltardiff}--\ref{eq:rgen}), 
supplemented by the correction,
Eq. (\ref{eq:pinchoffcorr}), and assuming  $\tau_s/\tau=4\cdot 10^{5}$. 
Tunnelling parameters $G_0/\nu_0$  in units $10^{-16} {\rm A} \cdot {\rm m}$ 
and $K/\nu_0^3$  in units $10^{-40}$ ${\rm A}\cdot {\rm m}^5$: 
for $U_G(1)=6V$:  $0.09$ 
and $0.007$;  
for $U_G(1)=7V$:  $0.5$ 
and $0.045$; 
for $U_G(1)=8V$:  $0.6$ 
and $0.1$; 
for $U_G(1)=9V$:  $1.6$ 
and $0.25$, and 
for $U_G(1)=12V$:  $1.7$ 
and $2.0$.
}
\end{figure*}

In Fig. \ref{fig:gatefit} we show the antisymmetric part of 
measured resistivity for different values of the gate voltage
$U_G(1)$, as plotted in Fig. \ref{fig:datagate}. When attempting 
to fit these curves theoretically by means of Eqs.
(\ref{eq:deltardiff}--\ref{eq:rgen}) (dashed lines in Fig. 
\ref{fig:gatefit}), we find that there is a considerable freedom in the choice
of the suitable parameter values. Indeed, there is no independent data 
on the values of the tunnelling parameters $G_0$ and
$K$ [see Eqs. (\ref{eq:tuncur}--\ref{eq:tuncon})], nor on their 
dependence on the carrier density $n$. While the value of
$K$ determines the saturation current density $j_{\rm sat}$ [see 
Eq. (\ref{eq:jsat})], and therefore the bending of the
theoretical curve for $R_A(I_{\rm DC})$, the effect of increasing $G_0$ 
is rather similar to that of decreasing $\tau_s$, hence
the values of the latter two parameters are not uniquely determined 
by the profile of an experimental curve. In the fit
shown in Fig. \ref{fig:gatefit} we assumed that the ratio 
$\tau_s/\tau$ does not depend on the carrier density and equals
$4\cdot 10^{5}$. The latter choice appears not unreasonable, 
as it is close to the values reported earlier
for Si/Si-Ge quantum wells\cite{Wilamowski,Tyryshkin} and for 
Si-MOSFETs\cite{Shankar}. It differs from the value we 
used in fitting Fig. \ref{fig:linearfit} above 
($\tau_s/\tau = 1.7 \cdot 10^{5}$) because presently we include an additional
mechanism (tunnelling to a ``spin reservoir'', with $G_0 \neq 0$). Note
that this adjustment of the value of $\tau_s/\tau$ is {\it not} an order
of magnitude change: the value we use here remains within the experimental
range (see discussion in Sect. \ref{sec:overview}).

\begin{figure}
\includegraphics[width=9.0cm]{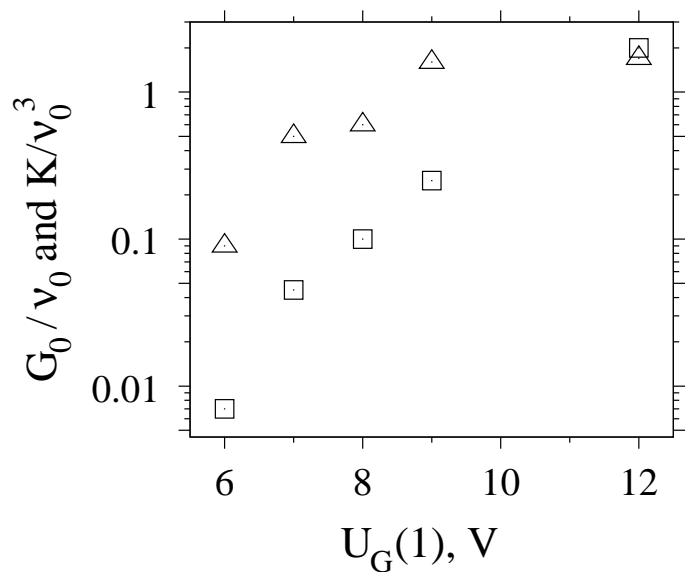}
\caption{\label{fig:tunpar}.
Tunnelling parameters  $G_0/\nu_0$  (triangles; in units $10^{-16} {\rm A} \cdot {\rm m}$)
and $K/\nu_0^3$  (squares; in units $10^{-40}$ ${\rm A}\cdot {\rm m}^5$)
for different values of the gate voltage  $U_G(1)$.
}
\end{figure}

The dependencies of the tunnelling parameters used in Fig. \ref{fig:gatefit} on
the gate voltage are shown in Fig. \ref{fig:tunpar}. As expected, both $G_0$ 
and $K$ increase with increasing $U_G(1)$, as  the barrier height
becomes lower relative to the Fermi energy. Indeed, the entire potential
energy landscape (including the tunnelling barriers) is pushed down in energy
by increasing gate voltage.  

Our assumption that the ratio $\tau_s/\tau$ 
is independent of $n$ was experimentally verified\cite{Wilamowski} 
for the case of the Si/Si-Ge quantum wells, where it is indeed an 
expected property of Dyakonov-Perel' spin relaxation mechanism.
The case of Si-MOSFETs might be different, but the experimental data on
the dependence of $\tau_s$ on $n$ and/or $\tau$ in a Si-MOSFET are lacking.
As explained above, in fitting the experimental data in Fig. \ref{fig:gatefit}
we could have used different values of 
$\tau_s/\tau$ for each $U_G(1)$; still, perfect fits would 
have been obtained by appropriately
choosing $G_0$ in each case.

We conclude that our theory appears capable of a perfect 
description of the measured antisymmetric part of the resistance.
A more definitive verification of our theoretical picture 
(and perhaps identification of the underlying microscopic mechanisms)
should be possible once the experimental values of the relevant 
system parameters become available.

\section{Conclusion}
\label{sec:conclu}

The  observed asymmetric behaviour of the resistance, $R(I_{\rm DC})$, 
of the 2DEG in a slot-gate Si-MOSFET in 
a parallel magnetic field is clearly due to the spin transport 
properties of this system. We suggested a phenomenological
model which features effective spin injection into the slot region; 
the rate of this spin injection is controlled by the DC
current.

Indeed, the area where the carrier density is smaller is characterised by
a stronger spin polarisation (cf. Fig \ref{fig:energyscheme}). Hence an
electron flow  (electrical current) in this area is accompanied by
a transfer of larger spin per unit time (spin current) than in the case of the
same electrical current flowing through the area with larger density.
It follows that when electrons flow from the area with smaller carrier density
into the area with higher density via the slot region, they must leave excess
spin in the vicinity of the slot (spin injection). This translates into
an increase of local magnetisation in this region (spin accumulation;
cf. Fig.\ref{fig:deltam}). This increase is, of course, not unlimited but 
rather moderated by spin relaxation and diffusion processes. Similarly, when 
the electrons flow in the opposite  direction, local magnetisation near the 
slot decreases (spin depletion).

The resultant deviation of local spin polarisation of the 2DEG from 
equilibrium affects the sample resistance, the linkage
being provided for the most part by the well-known positive  
magnetoresistance phenomenon. Since this phenomenon is of spin origin,
the resistivity depends on the magnetic field via magnetisation. An  
increase of the local magnetisation thus
leads to an increased resistivity in the region near the slot, and hence
to an increased overall resistance. Similarly, spin depletion near the slot
results in a decrease in resistance. 
Our theory yields a good quantitative description
of the resistance asymmetry for relatively small values of the DC current, 
where the antisymmetric part $R_A$ of the resistance is linear in 
$I_{\rm DC}$. 
Therefore it appears certain that we have captured the correct physical 
mechanism, providing an adequate
explanation for the resistance asymmetry in general.

At the same time, the dependence of $R_A$ on $I_{\rm DC}$ at stronger currents 
becomes sublinear ("saturation"). This implies
the presence of an additional, non-linear route for dissipating 
the non-equilibrium magnetisation density near the slot.
We suggest that this is due to  tunnelling into a "spin reservoir", 
which could be exemplified by the underlying $p$-type silicon,
although other options are also possible. With such a non-linear 
tunnelling added to our model, we are able to fit the
experimental curves for $R_A(I_{\rm DC})$ throughout the entire range 
of current values. This raises the problem of identifying the
precise nature of the "spin reservoir" and directly measuring the 
tunnelling parameters. In addition, systematic measurements
of the dependence of spin relaxation rate on carrier density in a 
conventional Si-MOSFET (without a slot in the gate) still have to be 
performed, providing
another important input parameter for our theory.

From a broader prospective, we describe and interpret an unusual 
magnetotransport phenomenon, taking place  in a 2DEG
with abruptly varying carrier density, in the presence of a parallel 
magnetic field. The specific realisation of this system
(slot-gate Si-MOSFET) can be viewed as incidental. The observed 
resistance asymmetry highlights new and interesting features
of low-dimensional spin and charge transport, and may point to 
additional possibilities for spin manipulation in microtechnology.
It is generally recognised that the efficiency of spintronic devices 
is limited by finite spin lifetime, due to usual spin-relaxation
mechanisms.  The observed saturation of resistance asymmetry at 
stronger current suggests that there are other, significantly 
non-linear, effects which may have to be taken into account.

\acknowledgments{The authors take pleasure in thanking R. Berkovits, 
I. V. Gornyi, B. D. Laikhtman, and L. D. Shvartsman for
discussions, and A. Belostotsky and A. Bogush  for assistance. 
This work was supported  by the  Erick and 
Sheila Samson Chair of Semiconductor Technology and by the Israeli 
Absorption Ministry.}


\begin{thebibliography}{99}
\bibitem{jcr} Careful examination shows that even if the values of 
$j_{{\rm cr},i}$ were much lower, the crossover from diffusion to drift regime
of Eq. (\ref{eq:spindd}), occurring at $j \sim j_{\rm cr}$, would not be 
sufficient
to explain the observed saturation of $R_A(I_{\rm DC})$ at large currents
(cf. Fig. \ref{fig:linearfit}).
\end{thebibliography}
\end{document}